\documentclass[a4paper,11pt]{article}
\usepackage{amsmath,amssymb}
\usepackage[top=3cm,bottom=3cm,left=3cm,right=3cm]{geometry}
\usepackage{color}
\usepackage[T1]{fontenc}
\usepackage{hyperref}
\hypersetup{colorlinks=true}
\usepackage{cite}
\date{}
\begin{document}
\author{Burak Tevfik Kaynak${^\dag}$ and O. Teoman Turgut$^\ddag$ \\ Department of Physics, Bo\u{g}azi\c{c}i University \\ 34342 Bebek, Istanbul, Turkey \\ $^\dag$burak.kaynak@boun.edu.tr, $^\ddag$turgutte@boun.edu.tr}
\title{\bf Infinitely many singular interactions on noncompact manifolds}
\maketitle
\begin{abstract}
We show that the ground state energy is bounded from below when there are infinitely many attractive delta function  potentials placed in arbitrary locations, while all being separated at least  by a minimum distance, on two dimensional non-compact manifold. To facilitate the reading of the paper, we first present the arguments in the setting of Cartan-Hadamard manifolds and then subsequently discuss the general case.  For this purpose, we employ the heat kernel techniques as well as some comparison theorems of Riemannian geometry, thus generalizing the arguments in the flat case following the approach presented in Albeverio et.  al. (2004).
\end{abstract}
\section{Introduction}\label{int}
In this article, we would like to revisit the study of point interactions in Riemannian manifolds, as presented in Ref.~\cite{et} to understand  the case of  infinitely many point centers on a non-compact manifold.  Since the problem by its nature is nonperturbative and is defined on a Riemannian manifold, the authors of Ref.~\cite{et}  study the model through the resolvent of the full Hamiltonian. The resolvent contains an additional part, expressed via the inverse of an operator, so-called principal operator, formulated  completely with the help of the heat kernel of the Laplace operator on the manifold (this form of the resolvent is known as Krein's formula in the mathematics literature). The use of  heat kernel of  the Laplacian within  the resolvent has a definite advantage for  addressing  both the geometry and the renormalization related issues on the same footing. This becomes especially transparent when one deals with singular interactions supported on curves and submanifolds, which received some attention recently for curves~\cite{c1,c3,c4,c5,c6,c7,c8,c9,c10} and surfaces~\cite{c8,s2,s3,s4,s6} embedded in flat space. When the ambient space becomes a curved manifold, using heat kernel techniques turns out to be essential with which regularization is naturally accomplished, as it is developed in Refs.~\cite{bt1,bt2,bt3}. Thus, it is, in these references, shown that this  method upon which the whole formalism is based,  allows one to work in a natural environment to analyze the interrelations between the geometry of the manifold, the bound-state energy, and the renormalization aspects of these models.

In this article our pursuit is  to extend the study of countably infinitely many point interactions in $\mathbb{R}^2$, as it is presented in Chapter 3  in the authoritative book~\cite{al}, to the case in which the underlying space is replaced by a general two dimensional noncompact manifold. An equally interesting and difficult problem appears in  the one dimensional case when the delta functions are all arbitrarily distributed with no minimum distance conditions  and with arbitrary  strengths. The spectral aspects of this problem and various conditions on the strengths and distances are thoroughly  investigated in  the recent work by Kostenko and Malamud~\cite{km}.  It is an interesting challenge to understand the spectral aspects of the present problem under more general assumptions for the coupling strengths. A more recent study on the spectral aspects of infinitely many point interactions in three dimensional flat space is studied in Ref.~\cite{mk}, where the reader can find further references on the subject.  Our main result can be summarized as follows: Consider two dimensional noncompact manifold whose sectional curvature is bounded from below. On such a manifold, we introduce an infinite number of point interactions on arbitrary locations. If all the bound state energies supported by these point centers are bounded from below by a common value $\mu_*$ and moreover all the distances between these points are bounded from below by a minimum distance $d_{\min}$, then one can compute a lower bound to the ground state energy, expressed in terms of $d_{\min},\mu_*$ and the geometric data of the manifold. This problem has importance especially when one thinks of these delta function centers as models of  impurities, appearing in random locations inside the manifold, and avoiding one another by a certain distance by considerations of energy.  It is certainly a very interesting and  a very challenging problem, which we plan to return in a subsequent work,  to consider singular interactions supported both on infinitely many curves or  submanifolds, all of which are embedded in some non-compact Riemannian manifolds, extending what is presented in Refs.~\cite{bt1,bt2,bt3}. Some results in this direction are obtained for special geometries in flat space in Ref.~\cite{akm}.

The plan of the article is as follows: in Section~\ref{cons}, the model will be constructed as a limit of well-defined projection operators through the aforementioned resolvent formula. Afterwards our choice of regularization of the principal operator will be shown to suffice to renormalize the model. In Section~\ref{pro}, the proof of which the ground-state energy is finite will be presented by exploiting some remarkable estimates of the heat kernel of Cartan-Hadamard  manifolds, using  the Holmgren bound for the norm of countably infinite matrices, and lastly relying on  the remarkable comparison theorem of Toponogov. In Section~\ref{non}, we present the proof for general noncompact manifolds, and emphasize the main changes.
It is implicit that the construction of the resolvent is also accomplished following the approach presented in \cite{al}, since this does not require any new ideas, we only present the details of lower bound for the ground state  energy.
\section{Construction of the renormalized Hamiltonian}\label{cons}
In this section, we review the construction of the resolvent operator on two dimensional manifolds corresponding to the Hamiltonian with countably infinitely many singular interactions, that are at least separated from each other by a global minimum distance.

Let us consider a Hamiltonian operator $H$ in $\mathcal{L}^2(\mathcal{M})$ with countably infinitely many singular interactions. A convenient way to express the interaction part of the Hamiltonian in a manifold is to write it as a projection operator. In this respect, we introduce the following family of functions,
\begin{align}
\Gamma_i^\epsilon(x) &= K_{\epsilon / 2} (x,p_i) \,,
\end{align}
$p_i$ being the points at which the interaction centers are located, and they are labeled by $i$. $K_t (x,y)$ is the heat kernel on the manifold. A natural interpretation of the parameter $\epsilon$ here is a cut-off due to its correspondence with a space cut-off. It is, therefore, going to be used as a regularization parameter in the forthcoming pages. The heat kernel is the positive fundamental solution of the heat equation for the Laplace operator $-\nabla_{g}^2$,
\begin{align}
-\frac{\hbar^2}{2 m} \nabla_g^2 K_t(x,x')  = - \hbar \frac{\partial}{\partial t} K_t(x,x') \,,
\end{align}
where the Laplace operator acts on the variable $x$. Its initial condition is given by $K_t(\cdot,\cdot) = \delta_g(\cdot,\cdot)$ as $t \rightarrow 0^+$, and it moreover enjoys the semi-group property,
\begin{align}
\int d \mu(x) K_u (x',x) K_v(x,x'') &= K_{u+v}(x',x'') \,,
\end{align}
where $d \mu(x)$ is the integration measure at the point $x \in \mathcal{M}$. We can write the inner product of these functions with the help of the semi-group property, and it gives
\begin{align}
\langle \Gamma_i^\epsilon \vert \Gamma_j^\epsilon \rangle &= K_\epsilon (p_i,p_j) \,.
\end{align}
The right-hand side of this expression is the generalized Dirac-Delta distribution in the limit $\epsilon \rightarrow 0^+$. The full Hamiltonian operator, in terms of both the free Hamiltonian and the projection operator constructed by the functions defined above, takes the following form,
\begin{align}
H^\epsilon &= - \frac{\hbar^2}{2 m} \nabla_g^2 - \sum_{i=1}^\infty \lambda_i(\epsilon) \vert \Gamma_i^\epsilon \rangle \langle \Gamma_i^\epsilon \vert \,,
\end{align}
where $m$ is the mass of the particle and $\lambda_i$ are the coupling strengths of each one of the interaction centers, labeled by $i$. Since the primary aim of this work is the study of the bound state structure of the model, the full resolvent corresponding to the Hamiltonian above is needed. This requires inverting the following equation,
\begin{align}
\vert \varphi \rangle &= \left( H^\epsilon - E \right) \vert \psi \rangle \,,
\end{align}
for any state $\vert \psi \rangle$ in the domain of $H^\epsilon$ for any given vector $\vert \varphi \rangle$ in the Hilbert space. Its calculation in terms of the resolvent of the free Hamiltonian is a standard procedure, and one can obtain the resolvent by solving the following equation for $\langle \tilde{\Gamma}_i^\epsilon \vert \psi \rangle$,
\begin{align}\label{psi}
\vert \psi \rangle &= \sum_i (H_0 - E)^{-1} \vert \tilde{\Gamma}_i^\epsilon \rangle \langle \tilde{\Gamma}_i^\epsilon \vert \psi \rangle + (H_0 - E)^{-1} \vert \varphi \rangle \,,
\end{align}
where the tilde $\tilde{}$ on the objects stands for $\Gamma_i^\epsilon$ being scaled by the coupling constants $\sqrt{\lambda_i}$. The solution of this equation is a matrix equation, satisfied by $\langle \tilde{\Gamma}_i^\epsilon \vert \psi \rangle$, and is given by
\begin{align}\label{gpsi}
\langle \tilde{\Gamma}_i^\epsilon \vert \psi \rangle &= \sum_j \left[ \frac{1}{1-\langle \tilde{\Gamma}^\epsilon \vert (H_0-E)^{-1} \vert \tilde{\Gamma}^\epsilon \rangle} \right]_{ij} \langle \tilde{\Gamma}_j^\epsilon \vert (H_0-E)^{-1} \vert \varphi \rangle \,.
\end{align}
Eqs.~(\ref{psi},\ref{gpsi}) result in the solution for the full resolvent we seek, and it is, after scaling all $\tilde{\Gamma}_i^\epsilon \vert \psi \rangle$ back, given by 
\begin{align}\label{fr}
\frac{1}{H^\epsilon-E} &= \frac{1}{H_0-E} + \frac{1}{H_0-E} \vert \Gamma_i^\epsilon \rangle \frac{1}{\Phi^\epsilon_{ij}} \langle \Gamma_j^\epsilon \vert \frac{1}{H_0-E} \,, 
\end{align}
in which summations are meant over the repeated indices. The operator $\Phi_{ij}^\epsilon$ refers to a matrix operator, which is called the principal operator and also known as the Krein function in the mathematics literature. The parameter $\epsilon$ as a superscript here indicates that the model may require a renormalization, and we will see that this is the case. The major importance of this operator stems from the fact that this matrix operator itself keeps all the information related to the bound state structure. It is the only operator that has its inverse in the full resolvent formula~(\ref{fr}) other than the free Hamiltonian. Since the poles of the full resolvent below the spectrum of the Laplacian corresponds to bounds states, the bound state spectrum must be determined by the eigenvalues of the principal operator below. The eigenvalues are monotonically decreasing functions of $E$ for $E \in \mathbb{R}$. The unique values of $E$ which makes a particular eigenvalue zero corresponds to the physical bound state energies, assuming that it is below the spectrum of the free Hamiltonian.

It can easily be shown that the principal operator is given by
\begin{align}
\Phi_{ij}^\epsilon(E) &= \left\{ \begin{array}{l} \displaystyle{\frac{1}{\lambda_i} - \langle \Gamma_i^\epsilon \vert (H_0 - E)^{-1} \vert \Gamma_i^\epsilon \rangle} \,, \quad i=j \,, \\ - \displaystyle{ \langle \Gamma_i^\epsilon \vert (H_0 - E)^{-1} \vert \Gamma_j^\epsilon \rangle} \,, \quad i \neq j \,. \end{array} \right.
\end{align}
In order to calculate explicitly the principal operator, we will rewrite the free resolvent as an integration of the heat kernel of the free Hamiltonian over its ``time'' parameter.
\begin{align}
\langle x \vert\frac{1}{H_0-E} \vert x' \rangle &= \int_0^\infty \frac{d t}{\hbar} \, e^{E t/ \hbar} K_t(x, x') \,.
\end{align}
With the help of both this integral representation and the semi-group property of the heat kernel we can easily obtain the principal operator. Let us first look at the diagonal part of the principal operator after doing those,
\begin{align}
\Phi_{ii}^\epsilon (E) &= \frac{1}{\lambda_i(\epsilon)} - \int_\epsilon^\infty \frac{d t}{\hbar} \, e^{E(t - \epsilon)  / \hbar} K_t(p_i,p_i) \,.
\end{align}
This expression is obviously divergent in the limit $\epsilon \rightarrow 0^+$. A natural choice to regularize this expression is to choose the coupling strength of each one of the interaction centers,
\begin{align}\label{lei}
\frac{1}{\lambda_i(\epsilon)} &= \int_\epsilon^\infty \frac{d t}{\hbar} \, e^{- \mu^2 (t-\epsilon) / \hbar} K_t(p_i,p_i) \,.
\end{align}
The limit $\epsilon \rightarrow 0^+$ is now safely taken after the choice~(\ref{lei}), and we reach the expression below for the diagonal part of the principal operator,
\begin{align}\label{pii}
\Phi_{ii}(E) &= \int_0^\infty \frac{dt}{\hbar} \left( e^{-\mu_i^2 t / \hbar} - e^{E t / \hbar} \right) K_t(p_i,p_i) \,.
\end{align}
Despite the fact that this expression is valid for $E<0$, its extension to the complex plane is possible by a proper analytic continuation.

Our choice of regularization has the following interpretation: If we isolate the center at location $p_i$, it has a bound state at energy $-\mu_i^2$. Other regularization prescriptions are also possible, which would lead to the same physical result as explained in Ref.~\cite{et}.
It is also convenient to choose $E = -\nu^2$ in search of bound state solutions, and it will be the case, henceforth. Thus, the full principal operator is now given by
\begin{align}\label{phi}
\Phi_{ij}(-\nu^2) &= \left\{ \begin{array}{l} \displaystyle{\int_0^\infty \frac{dt}{\hbar} \left( e^{-\mu_i^2 t / \hbar} - e^{-\nu^2 t / \hbar} \right) K_t(p_i,p_i)}  \,, \quad i=j \,, \\ - \displaystyle{\int_0^\infty \frac{d t}{\hbar} \, e^{-\nu^2 t / \hbar} K_t(p_i,p_j)} \,, \quad i \neq j \,. \end{array} \right.
\end{align}
\section{Proof of the finiteness of the ground-state energy for Cartan-Hadamard manifolds}\label{pro}
A Cartan-Hadamard manifold is a Riemannian manifold $(\mathcal{M},g)$ that is complete and simply-connected, and has everywhere non-positive sectional curvature as further explored from the point of heat kernels in Section 7.4 in Ref.~\cite{gri1}. In this study, we will, however, restrict ourselves to the family of two dimensional Cartan-Hadamard manifolds with sectional curvature satisfying $\mathrm{Sec} \geq - \kappa$ such that $\kappa \in \mathbb{R}^+$.

When there is a finite number of delta functions on a manifold,  it is possible to show, by elementary matrix analysis,  that the ground state energy remains bounded from below as long as the centers are separated by nonzero distances. This question becomes more subtle when we have an infinite number of such centers each one of which has finite strength, $\mu_i$. It is clear that in this case as well we must demand the infimum of the separations between centers to have a fixed nonzero value,  we call as  $d_{\min}$. With the intuition we gained from the flat space $\mathbb{R}^2$ case we expect that this should be sufficient to have a lower bound on the ground state energy. Indeed studying the two-delta function problem and letting the distance between them go to zero, one would see that the ground state energy diverges. This shows that the minimum distance requirement is necessary. We  aim to prove that this expectation is true  for the class of Cartan-Hadamard manifolds in two dimensions. The truth of this statement was already established in the flat space. The essential idea in this direction was  found in the flat space case by Albeverio et al as presented in Chapter 3 in Ref.~\cite{al}. The key step involves estimating the norm of the off-diagonal part  of the principal matrix in terms partial sums of the off-diagonal parts, which is given efficiently by the Holmgren bound as we will discuss below.  

To show that the energy is bounded from below we will employ a simple criteria. We will split the infinite matrix  $\Phi_{ij}$ into a diagonal and an  off-diagonal part, and show that above  a certain  value  of the energy  parameter  $\nu$, this matrix is always invertible, which means that below a certain energy $\Phi(E)$ matrix cannot have any zero eigenvalues, hence the ground state energy of the system is bounded from below. 
Note that we can write, 
\begin{align}
 \Phi &= D + O = D^{1/2} \left(1+D^{-1/2}OD^{-1/2} \right) D^{1/2} \,.
\end{align}
We recall that the matrix inside the parentheses will be invertible if the Neumann series of it converges. This implies that we should have 
\begin{align}
 \| D^{-1/2} O D^{-1/2} \| &<1 \,.
\end{align}
Let us remind that we set $E = -\nu^2$. With this choice one can see that the matrix elements of $O(\nu)$ are decreasing functions of $\nu$ and the ones of $D$ are increasing functions of $\nu$. Hence if we satisfy this norm bound for a specific value of $\nu$, it holds for all the values above this one. To make sure that there is a solution we will employ an upper bound for the above norm:
\begin{equation}
   \| D^{-1/2} O D^{-1/2} \| < \| D^{-1}\| \| O\| < \left({\rm sup}_i [D^{-1}_{ii}] \right) \| O \| \,.
\end{equation}

In this respect, we are in need of a lower estimate to the on-diagonal part of the principal operator~(\ref{phi}). In doing so, the Cheeger-Yau bound in Ref.~\cite{chow} will be of benefit to estimating the heat kernel from below. The theorem states that if $(\mathcal{M},g)$ is a complete Riemannian manifold with $\mathrm{Ric} \geq (n-1) K$ for some $K \in \mathbb{R}$, then the heat kernel satisfies
\begin{align}\label{yau}
K_t(x,x') &\geq K_t^K \left( d_g(x,x') \right) \,,
\end{align}
$K_t^K \left( d_g(x,x') \right)$ being the heat kernel on a simply connected, complete, $n$-dimensional Riemannian manifold of constant sectional curvature, which is also called a model space, and $d_g(x,x')$ here refers to the metric distance on the original manifold. In this study, we are only interested in a two dimensional problem, thereby interchangeably employing the lower bounds on Ricci and sectional curvatures. Since the manifolds involving in our problem have non-positive sectional curvatures, the heat kernel to be used on the right-hand side of the theorem above is the one of the two dimensional hyperbolic space. An explicit formula can be found in Ref.~\cite{dav}, and it is rewritten in terms of proper physical parameters with the correct dimensions, as
\begin{align}
K_t^\kappa(x,x') &= \frac{\sqrt{2} \kappa e^{- \frac{\hbar \kappa }{2 m} \frac{t}{4}}}{\left(  4 \pi \frac{\hbar \kappa}{2 m}t \right)^{3/2}} \int_{\sqrt{\kappa}d(x,x')}^\infty d \xi \, \frac{\xi e^{-  \frac{2 m}{\hbar \kappa}  \frac{\xi^2}{4 t}}}{\sqrt{\cosh \xi - \cosh \left(\sqrt{\kappa} d(x,x')\right)}}\,.
\end{align}     
That the on-diagonal part is of interest simplifies this heat kernel in such a way that we can get rid off the square root in the denominator. After some trigonometric and algebraic manipulations, the on-diagonal part of the principal operator is given by
\begin{align}
\Phi_{ii}(-\nu^2) &\geq \sqrt{\frac{2}{\kappa}} \left( \frac{m}{\pi \hbar} \right)^{3/2} \int_0^\infty d \xi \frac{\xi}{\sinh \xi} \int_0^\infty \frac{d t}{\hbar} \frac{e^{- \frac{2 m \xi^2}{\hbar \kappa t}}}{t^{3/2}} \left[ e^{-\left( \mu_i^2 + \frac{\hbar^2 \kappa}{8 m} \right) \frac{t}{\hbar}} - e^{-\left( \nu^2 + \frac{\hbar^2 \kappa}{8 m} \right) \frac{t}{\hbar}} \right] \,.
\end{align}
The integration over the time parameter results in
\begin{align}\label{pxi}
\Phi_{ii}(-\nu^2) &\geq \frac{m}{\pi \hbar^2} \int_0^\infty d \xi \, \frac{e^{-\sqrt{1 + \frac{8 m}{\hbar^2 \kappa} \mu_i^2} \xi} - e^{-\sqrt{1 + \frac{8 m}{\hbar^2 \kappa} \nu^2} \xi}}{\sinh \xi} \,.
\end{align} 
Due to the convexity of the function $\sinh$, we can here employ the Hermite-Hadamard inequality in Ref.~\cite{mit} so as to lessen the right-hand side. The Hermite-Hadamard inequality gives rise to the following inequality, satisfied by the function $\sinh$,
\begin{align}\label{sl}
\sinh(\xi) &\leq \frac{\xi}{2} (1 + \cosh \xi) \quad \text{for }\xi \geq 0\,.
\end{align}
Moreover, we can even enlarge the right-hand side of this inequality~(\ref{sl}) by utilizing $1 + \cosh \xi = 2 \cosh^2 (\xi /2)$ and $\cosh \xi \leq e^\xi$ for $\xi \geq 0$ in succession. Therefore, the inequality~(\ref{pxi}) transforms into
\begin{align}
\Phi_{ii}(-\nu^2) &\geq \frac{m}{\pi \hbar^2} \int_0^\infty d \xi \, \frac{e^{-\left( 1 + \sqrt{1 + \frac{8 m}{\hbar^2 \kappa} \mu_i^2} \right)\xi} - e^{-\left( 1 + \sqrt{1 + \frac{8 m}{\hbar^2 \kappa} \nu^2} \right)\xi}}{\xi} \,.
\end{align}
The integral above is none other than the integral representation of a difference of two logarithms. Thus, the lower estimate of the on-diagonal part of the principal operator is given by
\begin{align}\label{plb}
\Phi_{ii}(-\nu^2) &\geq \frac{m}{\pi \hbar^2} \log \left( \frac{1 + \sqrt{1 + \frac{8 m}{\hbar^2 \kappa} \nu^2}}{1 + \sqrt{1 + \frac{8 m}{\hbar^2 \kappa} \mu_i^2}} \right) \,.
\end{align} 
It is obvious that the on-diagonal lower estimate of the principal operator for the flat space is recovered if one takes the limit $\kappa \rightarrow 0$, and it is given by
\begin{align}
\Phi_{ii}(-\nu^2) &\geq \frac{m}{2 \pi \hbar^2} \log \left(  \frac{\nu^2}{\mu_i^2} \right) \,.
\end{align}

Before we estimate the norm of the off-diagonal part of the principal operator, a natural method so as to count the numbers of the interaction points throughout the manifold is to be devised.  In pursuit of this, we, level by level, will determine the uniform arrangements of the surrounding interaction points that are spreading outwards from an arbitrarily chosen one among those special points of the manifold. Since this point is located at the center of this configuration, we will henceforth call this point a center. Each level, labeled by $l$, consists of surrounding interaction points that are not only $l d_{\min}$ away from the center but also $d_{\min}$ away from each other. Therefore, we have $n(l)$ numbers of petals in each level. It is obvious that every two petals in a level with the center form a unique geodesic triangle as a corollary of the Cartan-Hadamard theorem. The geodesics connecting each of the adjacent petals to the center form an angle, and the number
  of the petals should, apparently, be determined by the integer part of the ratio of $2 \pi$ to this angle.

Unfortunately, there is not any feasible way to compute this number in a general Cartan-Hadamard manifold. We can, however, estimate $n(l)$ from above instead. This will suffice for our purposes as it will be shown below. In order to achieve this aim, we will employ a very powerful comparison theorem of great importance in Riemannian geometry, the so-called Toponogov's theorem in Refs.~\cite{chern, cheeger}. Although there are different versions of the theorem, namely secant, angle and hinge comparisons, the angle comparison version thereof is what we need here. 

Let $\mathcal{M}$ be a two dimensional complete Riemannian manifold, and $\mathcal{M}_\delta$ be a two dimensional simply connected Riemannian manifold of constant sectional curvature $\delta$. Let $\triangle$ be a geodesic triangle in $\mathcal{M}$ formed by three arc length parametrized geodesic segments $\gamma_i$, $i=1,2,3$, of lengths $\vert \gamma_i \vert$ such that $\gamma_i \left(\vert \gamma_i \vert \right) = \gamma_{i+1}(0)$. If $\delta > 0$, then there is a side restriction due to the theorem by Myers and Bonnet in Ref.~\cite{cheeger}. However $\delta<0$ for Cartan-Hadamard manifolds, whereby we will not elaborate further upon this case. The angles between the geodesic segments $\alpha_i$ are defined by
\begin{align}
\alpha_i &= \sphericalangle \left( -\dot{\gamma}_{i+1} \left( \vert \gamma_{i+1} \vert \right) ,\dot{\gamma} _{i+2} ( 0 ) \right) \,,
\end{align}
such that $0 \leq \alpha_i \leq \pi$. The angle comparison version of this remarkable theorem in Ref.~\cite{chern} states that if $\mathcal{M}$ is  a complete Riemannian manifold with sectional curvature satisfying $\mathrm{Sec} \geq \delta$ such that $\delta \in \mathbb{R}$, there exists a comparison triangle $\triangle_\kappa$, also called an Aleksandrov triangle, in the model space $\mathcal{M}_\delta$ with the corresponding geodesics $\gamma_{\delta i}$ of lengths $\vert \gamma_{\delta i} \vert = \vert \gamma_i \vert$ such that the angles of $\triangle$ and $\triangle_\delta$ satisfy
\begin{align}
\alpha_{\delta i} & \leq \alpha_i \quad \text{for } i = 1,2,3 \,.
\end{align}

Now, we can estimate $n(l)$ from above by its value calculated in the model space. Since two dimensional Cartan-Hadamard manifolds with the sectional curvature satisfying $\mathrm{Sec} \geq - \kappa$ such that $\kappa \in \mathbb{R}^+$ only involve in this study, the model space to be considered here should be a two dimensional hyperbolic space $\mathcal{H}_\kappa$ as the one which we have used in the Cheeger-Yau bound in search of a lower estimate for the on-diagonal part of the principal operator. 
\begin{align}
n(l) &\leq n_\kappa (l) = \frac{2 \pi}{\alpha_\kappa} \,.
\end{align}
The angle $\alpha_\kappa$, formed by the geodesics emanating from the center, of the geodesic triangle $\triangle_\kappa$ whose sides are $l d_{\min}, l d_{\min}, d_{\min}$ in length follows the hyperbolic laws of cosines,
\begin{align}
\cos \alpha_\kappa &= \frac{\cosh^2 \left( \sqrt{\kappa} d_{\min} l\right) - \cosh \left( \sqrt{\kappa} d_{\min} \right)}{\sinh^2 \left( \sqrt{\kappa} d_{\min} l\right)} \,.
\end{align}
Since $1 - \cos \alpha = 2 \sin^2 \left( \alpha/2\right)$, the inequality transforms into
\begin{align}\label{n1}
n(l) &\leq \frac{\pi}{\arcsin \left[ \frac{\sinh \left( \sqrt{\kappa} d_{\min}/2 \right)}{\sinh \left( \sqrt{\kappa} d_{\min} l \right)} \right]} \,.
\end{align}
The limit $\kappa \rightarrow 0$ is here worthy of remark. Taking this limit allows us to obtain the number of the interaction points around a fixed interaction center in $\mathbb{R}^2$. For example, it results in $n(1) = 6$ for the first level, that is equal to the first centered hexagonal number if we exclude the center. This is the number of the circles in the first level for the hexagonal packing of circles around a circle.

Moreover, we can still go further in estimating the right-hand side of the inequality~(\ref{n1}) by first replacing $1/\arcsin(\xi)$ by $1/\xi$, and then playing with the hyperbolic functions in both the numerator and the denominator. After the replacement, the numerator can be enlarged whilst the denominator can be lessened through the Hermite-Hadamard inequality as being done before. One can easily obtain the following inequalities for the numerator and the denominator, respectively,
\begin{align}
\sinh \left( \sqrt{\kappa} d_{\min} l \right) &\leq \frac{\sqrt{\kappa} d_{\min} l}{2} \left[ 1 + \cosh \left( \sqrt{\kappa} d_{\min} l \right) \right] \,, \\
\sinh \left( \frac{\sqrt{\kappa} d_{\min}}{2} \right) &\geq \frac{\sqrt{\kappa} d_{\min}}{2} \cosh \left( \frac{\sqrt{\kappa} d_{\min}}{4} \right) \,.
\end{align}
After inserting the inequalities above into the inequality~(\ref{n1}), one has
\begin{align}
n(l) &\leq \pi l \frac{1 + \cosh \left( \sqrt{\kappa} d_{\min} l \right)}{\cosh \left( \frac{\sqrt{\kappa} d_{\min}}{4} \right)} \,.
\end{align}
The fact that $1 + \cosh(\xi) = 2 \cosh^2(\xi/2)$ and $\cosh(\xi) \leq e^\xi$ for $\xi \geq 0$ makes the right-hand side of the expression above even larger, whereby we obtain even a simpler upper bound. Thus, the number of the interaction points around a fixed center is estimated from above by the following expression, 
\begin{align}\label{nl}
n(l) & \leq 2 \pi l \mathrm{sech} \left(\frac{\sqrt{\kappa} d_{\min}}{4} \right) e^{\sqrt{\kappa} d_{\min} l} \,.
\end{align}

We now define $\sup_i\mu_i=\mu_* < \infty$. Using the lower bound obtained in the inequality~(\ref{plb}) for the diagonal elements, we find a uniform lower bound, 
 \begin{align}\label{dia}
\Phi_{ii}(-\nu^2) &\geq \frac{m}{\pi \hbar^2} \log \left( \frac{1 + \sqrt{1 + \frac{8 m}{\hbar^2 \kappa} \nu^2}}{1 + \sqrt{1 + \frac{8 m}{\hbar^2 \kappa} \mu_i^2}} \right)\geq
  \frac{m}{\pi \hbar^2} \log \left( \frac{1 + \sqrt{1 + \frac{8 m}{\hbar^2 \kappa} \nu^2}}{1 + \sqrt{1 + \frac{8 m}{\hbar^2 \kappa} \mu_*^2}} \right)  \,.
\end{align} 
As a result, this leads to
\begin{align}
   \| D^{-1/2} O D^{-1/2} \|  &<
\frac{\pi \hbar^2}{m}\left[ \log \left( \frac{1 + \sqrt{1 + \frac{8 m}{\hbar^2 \kappa} \nu^2}}{1 + \sqrt{1 + \frac{8 m}{\hbar^2 \kappa} \mu_*^2}} \right)\right]^{-1} \| O \|  \,.
\end{align}
To further estimate $\| O \|$ we will use the well-known Holmgren bound, which is given in Lemma C.3 in Ref.~\cite{al},
\begin{align}
\| O \| &\leq \left[\sup_i \sum_{j,j\neq i} |O_{ij}| \sup_j \sum_{i,i\neq j} |O_{ij}| \right]^{1/2} \,.
\end{align}
Since the matrix $O$ is symmetric the two sums are identical and we have a simpler expression to work with. We now estimate the sums above, the first thing we note is the following upper bound of heat kernels on two dimensional Cartan-Hadamard manifolds, given in the expression~(7.15) with the explanation under Theorem 7.5 in Ref.~\cite{gri1} and in the expression~(15.49) in Ref.~\cite{gri2}, 
\begin{align}\label{chu}
  K_t(p_i,p_j) &\leq \frac{A}{4 \pi \frac{\hbar}{2 m}t}e^{-\frac{2 m d^2(p_i,p_j)}{B \hbar t}} \,,
\end{align}
in which $A$ and $B$ are some constants with $B$ to be chosen strictly larger than 4. If we write $B = 4 + \epsilon$ such that $\epsilon > 0$, then $A$ depends on the dimension of the manifold, the parameter $\epsilon$, and the isoperimetric function of the manifold with an isoperimetric constant specific to the manifold in question.
 
Employing the upper bound~(\ref{chu}) in the Holmgren bound produces an upper bound irrespective of how the centers are located. This bound is a decreasing function of the distances between the centers. In order to obtain the most stringent bound possible for the energy we must, therefore,  make this expression as large as possible. This can be achieved by choosing configurations which will make the mutual distances as small as possible. This cannot be obtained exactly, however, the following counting argument provides even a larger estimate. Let us assume that the index $j$ is fixed. We organize the sum corresponding to the fixed index $j$ according to the distance from the point $p_j$. The worse that could happen at a fixed distance away from the point $p_j$ is a tight packing of centers with all of them essentially situated as close as possible to the minimum distances with their nearest neighbors. An upper bound to the number of this densely packed neighbors is estimated in the inequality~(\ref{nl}). Then the sum of off-diagonal terms in the Holmgren bound can be bounded, as follows 
\begin{align}
\sum_{i\neq j} |O_{ij}| &\leq  \frac{A m}{\pi \hbar^2} \sum_{l=1}^\infty  n(l) K_0 \left(  2 \sqrt{\frac{2 m d_{\min}^2}{B \hbar^2}} l \nu\right) \,.
\end{align}
If we now employ the inequality, 
\begin{align}
K_0(x) &\leq \frac{2}{x} e^{- \frac{x}{2}} \,,
\end{align}
which is easily found from the integral representation 8.432.1 in Ref.~\cite{gra} and the upper bound on $n(l)$, given by the inequality~(\ref{nl}), we obtain 
\begin{align}
 \| O \| & \leq \frac{2 A m}{ \hbar^2 \nu} \sqrt{\frac{B \hbar^2}{2 m d_{\min}^2}} \mathrm{sech} \left(\frac{\sqrt{\kappa} d_{\min}}{4} \right) \sum_{l=1}^\infty e^{-l \sqrt{\frac{2 m d_{\min}^2}{B \hbar^2}} \left( \nu - \sqrt{\frac{B \hbar^2 \kappa}{2 m}}\right)} \,.
\end{align}
This can further be estimated, by assuming sufficiently large $\nu$ for convergence, 
as follows
\begin{align}
 \| O\| & \leq \frac{2 A m}{ \hbar^2 \nu} \sqrt{\frac{B \hbar^2}{2 m d_{\min}^2}} \mathrm{sech} \left(\frac{\sqrt{\kappa} d_{\min}}{4} \right) \int_0^\infty d l e^{-l \sqrt{\frac{2 m d_{\min}^2}{B \hbar^2}} \left( \nu - \sqrt{\frac{B \hbar^2 \kappa}{2 m}}\right)} \nonumber \\
 &= \frac{2 A B m}{\hbar^2} \frac{\hbar^2}{2 m d_{\min}^2} \frac{\mathrm{sech} \left(\frac{\sqrt{\kappa} d_{\min}}{4} \right)}{\nu \left( \nu - \sqrt{\frac{B \hbar^2 \kappa}{2 m}} \right)} \,.
\end{align}
There is obviously a finite value $\nu_*$ which would satisfy 
\begin{align}\label{neu}
2 \pi A B \frac{\hbar^2}{2 m d_{\min}^2} \frac{\mathrm{sech} \left(\frac{\sqrt{\kappa} d_{\min}}{4} \right)}{\nu_* \left( \nu_* - \sqrt{\frac{B \hbar^2 \kappa}{2 m}} \right)} &< \log \left( \frac{1 + \sqrt{1 + \frac{8 m}{\hbar^2 \kappa} \nu_*^2}}{1 + \sqrt{1 + \frac{8 m}{\hbar^2 \kappa} \mu_*^2}} \right) \,.
\end{align}
This is the desired bound to make the Neumann series convergent. This would imply that the infinite matrix $\Phi$ is invertible, hence none of its eigenvalues can be zero. The above value $\nu_*$ provides a lower bound for $E_{gr}$ since $E_{gr} > - \nu_*^2$, and $\nu_*$ is expressible purely in terms of $d_{\min}$, $\kappa$, and the geometric data hidden in the coefficients $A$ and $B$. 
\section{Generalization to generic noncompact manifolds}\label{non}
In this section we will generalize the result of Section~\ref{pro} into the case where the underlying Riemannian manifold is generally noncompact. The proof basically follows the same steps. Due to the lower bound~(\ref{yau}) the off-diagonal terms grow as the distance between the centers shrink. Thus, to get a lower bound for the energy we should look for configurations as densely packed as possible. The only differences, however, rely on the heat kernel estimate and the counting argument so as to estimate the off-diagonal part of the principal operator. The required off-diagonal upper bound on the heat kernel~\cite{gri3} in terms of proper physical parameters is given by
\begin{align}\label{upno}
K_t(p_i,p_j) &\leq \frac{\mathrm{const}}{4 \pi \min \left( \frac{\hbar t}{2 m}, \rho^2 \right)} \left[ 1 + \frac{2 m d^2(p_i,p_j)}{\hbar t}\right]^2 \exp\left( -\frac{\lambda t}{\hbar} - \frac{2 m d^2(p_i,p_j)}{4 \hbar t}\right) \,,
\end{align}
where $\lambda$ is the spectral gap, and $\rho$ is the radius of the geodesic ball. The latter is chosen in a way so as to ensure that the exponential map is a local diffeomorphism. Generally, $\rho$ is related to the local sectional curvature. It is, through this estimate, meant that local regions with positive curvatures are also allowed, such as local bumps. The sum of off-diagonal terms in the Holmgren bound is bounded as
\begin{align}\label{oij}
\sum_{i \neq j} \vert O_{ij} \vert &\leq \sum_{l=1}^\infty n(l) \int_0^{2 m \rho^2/\hbar} \frac{dt}{\hbar}\frac{C}{4 \pi \frac{\hbar t }{2 m}} \left( 1 + \frac{2 m d_*^2 l^2}{\hbar t}\right)^2 e^{-\frac{\nu^2 t}{\hbar} - \frac{2 m d_*^2 l^2}{4 \hbar t}} \nonumber \\
& \quad + \sum_{l=1}^\infty n(l) \int_{2 m \rho^2/\hbar}^\infty \frac{dt}{\hbar}\frac{D}{4 \pi \rho^2} \left( 1 + \frac{2 m d_*^2 l^2}{\hbar t}\right)^2 e^{-\frac{\nu^2 t}{\hbar} - \frac{2 m d_*^2 l^2}{4 \hbar t}} \,,
\end{align}
where we choose $\lambda =0$, and $d_* = d_{\min} =\min_{i,j} d(p_i,p_j)$. We can extend the domain of the integrals to $(0,\infty)$. The $t$-integrations, then, yield
\begin{align}\label{tint}
& \frac{C m}{\pi \hbar^2} \left[ \left( 1 + \frac{8 m d_*^2 \nu^2 l^2}{\hbar^2} \right) K_0\left(\sqrt{\frac{2 m d_*^2 \nu^2}{\hbar^2}} l\right) + 12 \sqrt{\frac{2 m d_*^2 \nu^2}{\hbar^2}} l K_1\left( \sqrt{\frac{2 m d_*^2 \nu^2}{\hbar^2}} l \right) \right] \nonumber \\
& + \frac{D}{4 \pi \rho^2 \nu^2} \sqrt{\frac{2 m d_*^2 \nu^2}{\hbar^2}} l \left(1 +  \frac{8 m d_*^2 \nu^2 l^2}{\hbar^2} \right) K_1\left(\sqrt{\frac{2 m d_*^2 \nu^2}{\hbar^2}} l\right) \nonumber \\
& + \frac{D}{4 \pi \rho^2 \nu^2}\frac{8 m d_*^2 \nu^2 l^2 }{\hbar^2} K_0\left(\sqrt{\frac{2 m d_*^2 \nu^2}{\hbar^2}} l\right) \,.
\end{align}
We will insert the following upper bounds on the modified Bessel functions of the second kind into the above result, which makes the right hand-side of the inequality~(\ref{oij}) even larger,
\begin{align}\label{k01}
K_0(\xi) &\leq K_{1/2}(\xi) = \sqrt{\frac{\pi}{2 \xi}} e^{-\xi} \,, \\
K_1(\xi) &\leq \left( 1 + \frac{1}{\xi} \right) e^{-\xi} \,.
\end{align}
The second inequality above can easily be achieved by the integral representation 8.432.2 in Ref.~\cite{gra}. We will not give the result of this replacement for the sake of brevity, however.

So far, we have mentioned nothing about the counting argument to be used in the above inequality that we found. Although it is intractable to obtain an upper bound on the count of the singular sites in a generic noncompact manifold as efficient as we did in the Cartan-Hadamard case, we can still give a reasonable upper bound here as well. The counting argument this time reads as follows: we, first, fix an interaction point in the manifold as a center, and then count the number of the surrounding petals, which are reached by geodesics, as we did in the previous section. This gives the number of the interaction points in the first level, namely $n(1)$. Since the heat kernel upper bound~(\ref{upno}) also allows local regions with positive curvature, we can count the interaction points within a geodesic ball, if this is the case, until we reach a region with negative curvature. Secondly, we repeat this operation on each petal in each level. Let us assume that there is a global maximum value of $n(1)$ all over the manifold, i.e. $n^* = \max_i n_i(1)$. If we use this maximum value for each interaction site, then we can achieve an upper bound of $n(l)$. That is to say, we employ an exact replica of the cell with the maximum number of petals in the first level on each interaction point so as to evaluate the number $n(l)$. Therefore, this counting argument results in the following over-estimate,
\begin{align}\label{nln}
n(l) &\leq e^{l \log n^*} \,.
\end{align}
Now, it is time to amalgamate the result of the $t$-integrations, the inequalities obtained for the Bessel functions, and the upper bound of the number of the interaction points in each level $n(l)$ into one in order to estimate further the right hand-side of the inequality~(\ref{oij}), i.e. to estimate $\| O \|$. We can \emph{replace the summation by an integration over the variable $l$ from $0$ to $\infty$}. Thus, the required upper bound after straightforward calculations takes the following form,
\begin{align}\label{onon}
\| O \| &\leq \frac{m}{\hbar^2} \frac{C}{\sqrt{2}} \left( \frac{\hbar^2}{2 m d_*^2 \nu^2} \right)^{1/4} \frac{1}{\left( \sqrt\frac{2 m d_*^2 \nu^2}{\hbar^2} - \log n^*\right)^{1/2}} \nonumber \\
& \quad + \frac{m}{\hbar^2} \left[12 C + \frac{D}{2 \pi} \left( \frac{\hbar^2}{2 m \rho^2 \nu^2} \right)\right] \frac{\left( 2 \sqrt\frac{2 m d_*^2 \nu^2}{\hbar^2} - \log n^*\right)}{\left( \sqrt\frac{2 m d_*^2 \nu^2}{\hbar^2} - \log n^*\right)^2} \nonumber \\
& \quad + \frac{m}{\hbar^2} \frac{3}{\sqrt{2}}\left( \frac{2 m d_*^2 \nu^2}{\hbar^2} \right)^{3/4} \left[C + \frac{D}{2} \left( \frac{\hbar^2}{2 m \rho^2 \nu^2}\right) \right] \frac{1}{\left( \sqrt\frac{2 m d_*^2 \nu^2}{\hbar^2} - \log n^*\right)^{5/2}} \nonumber \\
& \quad + \frac{m}{\hbar^2} \frac{4 D}{\pi} \left( \frac{2m d_*^2}{\hbar^2}\right) \left( \frac{\hbar^2}{2 m \rho^2}\right)\frac{\left( 4 \sqrt\frac{2 m d_*^2 \nu^2}{\hbar^2} - \log n^*\right)}{\left( \sqrt\frac{2 m d_*^2 \nu^2}{\hbar^2} - \log n^*\right)^4} \,,
\end{align}
which is valid as long as the following condition is fulfilled,
\begin{align}\label{logn}
\frac{2 m d_*^2\nu^2}{\hbar^2} &\geq \log^2 n^*\,.
\end{align}
Unlike the upper bound of the off-diagonal part of the principal operator, the estimate of the diagonal part of the principal operator~(\ref{dia}), obtained in the last section, is still valid. Thus, we have the same inequality here as in~(\ref{neu}) except the left hand-side is determined by the inequality~(\ref{onon}), and it reads as follows
\begin{align}\label{neunon}
& \frac{C}{\sqrt{2}} \left( \frac{\hbar^2}{2 m d_*^2 \nu_*^2} \right)^{1/4} \frac{1}{\left( \sqrt\frac{2 m d_*^2 \nu_*^2}{\hbar^2} - \log n^*\right)^{1/2}} \nonumber \\
& \quad + \left[12 C + \frac{D}{2 \pi} \left( \frac{\hbar^2}{2 m \rho^2 \nu_*^2} \right)\right] \frac{\left( 2 \sqrt\frac{2 m d_*^2 \nu_*^2}{\hbar^2} - \log n^*\right)}{\left( \sqrt\frac{2 m d_*^2 \nu_*^2}{\hbar^2} - \log n^*\right)^2} \nonumber \\
& \quad + \frac{3}{\sqrt{2}}\left( \frac{2 m d_*^2 \nu_*^2}{\hbar^2} \right)^{3/4} \left[C + \frac{D}{2} \left( \frac{\hbar^2}{2 m \rho^2 \nu_*^2}\right) \right] \frac{1}{\left( \sqrt\frac{2 m d_*^2 \nu_*^2}{\hbar^2} - \log n^*\right)^{5/2}} \nonumber \\
& \quad +  \frac{4 D}{\pi} \left( \frac{2m d_*^2}{\hbar^2}\right) \left( \frac{\hbar^2}{2 m \rho^2}\right)\frac{\left( 4 \sqrt\frac{2 m d_*^2 \nu_*^2}{\hbar^2} - \log n^*\right)}{\left( \sqrt\frac{2 m d_*^2 \nu_*^2}{\hbar^2} - \log n^*\right)^4} < \frac{1}{\pi} \log \left( \frac{1 + \sqrt{1 + \frac{8 m}{\hbar^2 \kappa} \nu_*^2}}{1 + \sqrt{1 + \frac{8 m}{\hbar^2 \kappa} \mu_*^2}} \right)
\end{align}
There exists a finite value $\nu_*$ that satisfies the inequality~(\ref{neunon}) as well as fulfilling the condition~(\ref{logn}). This implies the convergence of the Neumann series, hence ensuring that the ground-state energy is finite.   
\section{Acknowledgment}
O. T. Turgut would like to thank Meltem \"Unel for discussion at the initial stages of this work.
This work is supported by Bo\u{g}azi\c{c}i University BAP Project \#:6513.

\end{document}